\documentclass[11pt]{article}
\usepackage{fullpage}
\usepackage{cite}
\usepackage{graphicx}
\usepackage{amsmath}
\usepackage{amssymb}
\title{The gravitational cusp anomalous dimension from Newton's Law}
\date{}
\renewcommand{\vec}[1]{\mbox{\boldmath$ #1 $}}

\begin{document}
\bibliographystyle{utphys}
\newcommand{\msbar}{\ensuremath{\overline{\text{MS}}}}
\newcommand{\DIS}{\ensuremath{\text{DIS}}}
\newcommand{\abar}{\ensuremath{\bar{\alpha}_S}}
\newcommand{\bb}{\ensuremath{\bar{\beta}_0}}
\newcommand{\rc}{\ensuremath{r_{\text{cut}}}}
\newcommand{\Nd}{\ensuremath{N_{\text{d.o.f.}}}}
\setlength{\parindent}{0pt}

\titlepage
%\begin{flushright}
%Preprint numbers here \\
%\end{flushright}

\vspace*{0.5cm}

\begin{center}
{\Large \bf The gravitational cusp anomalous dimension from AdS space}

\vspace*{1cm}
\textsc{D. J. Miller\footnote{David.J.Miller@glasgow.ac.uk} and C. D. White\footnote{Christopher.White@glasgow.ac.uk} } \\

\vspace*{0.5cm} SUPA, School of Physics and Astronomy, University of Glasgow,\\ Glasgow G12 8QQ, Scotland, UK\\

\end{center}

\vspace*{0.5cm}

\begin{abstract}
Recently a new picture has been developed for examining Wilson lines, and the
corresponding anomalous dimensions which govern their renormalization 
properties. By making a particular coordinate transform, the calculation of
the cusp anomalous dimension in QED or QCD can be related to the energy of a 
pair of static charges in Euclidean Anti-de-Sitter (AdS) space. This paper 
shows how the same picture can be used to describe Wilson 
lines in quantum gravity. We show how the relevant cusp 
anomalous dimension (which has recently been shown to be one loop exact) can be
obtained using the Newtonian limit of General Relativity. We also show how both
the QED and gravity cases emerge as special cases of a general formulation,
and that a continuous parameter exists which interpolates between them. 
The results may be useful in examining the relations between gauge and gravity
theories.
\end{abstract}

\vspace*{0.5cm}

\section{Introduction}
Wilson lines have been studied for many years in a variety of 
contexts~\cite{Arefeva:1980zd,Polyakov:1980ca,Dotsenko:1979wb,Brandt:1981kf,
Korchemsky:1985xj,Ivanov:1985np,Korchemsky:1987wg,Korchemsky:1988hd,
Korchemsky:1988si,Basso:2007wd}, in both Abelian and non-Abelian gauge 
theories. In particular, they govern the structure of large logarithms in 
perturbation theory due to soft gluon emission~\cite{Korchemsky:1992xv,
Korchemsky:1993uz} as is also seen in other approaches such as 
soft collinear effective theory (SCET)~\cite{Bauer:2000ew,Bauer:2000yr,
Bauer:2001ct,Bauer:2001yt,Bauer:2002nz,Becher:2006nr,Becher:2006mr,
Becher:2007ty}, factorisation theorems~\cite{Yennie:1961ad,Sterman:1986aj,
Catani:1989ne}, or the path integral technique of~\cite{Laenen:2008gt}. 
Essentially, hard partons emitting soft gluons cannot recoil and thus can only
change by a phase.
For this phase to have the right gauge transformation properties to slot into
an amplitude, it must be a Wilson line evaluated along a contour given by the
hard momentum of the outgoing particle. The ultraviolet renormalisation 
properties of Wilson lines govern the infrared singularities of scattering 
amplitudes~\cite{Ivanov:1985np,Korchemsky:1985xj,Korchemsky:1985xu,
Korchemsky:1985ts,Korchemsky:1986fj,Korchemsky:1987wg,Korchemsky:1991zp}, 
a fact which has recently been used to derive an all-order ansatz
for the infrared singularities of QCD scattering amplitudes~\cite{Dixon:2008gr,
Gardi:2009qi,Gardi:2009zv,Becher:2009cu,Becher:2009qa}, whose consequences
have been discussed further in~\cite{Becher:2009kw,Dixon:2009ur,DelDuca:2011ae,
DelDuca:2011xm}. More exotic applications of Wilson lines come from the 
interface between gauge and string theory, where it has recently been 
conjectured that certain Wilson loop configurations are dual to scattering 
amplitudes in ${\cal N}=4$ Super-Yang Mills theory at strong (and weak)
coupling~\cite{Alday:2007hr,Drummond:2007aua}. \\

Correlators of Wilson line operators have ultraviolet (UV) divergences. For 
smooth contours, these are completely removed by renormalisation of the gauge 
theory coupling. For contours which contain cusps or intersections, however, 
additional UV singularities are present which, after renormalisation of the 
coupling, factor out into an overall multiplicative 
factor~\cite{Dotsenko:1979wb,Brandt:1981kf,Korchemsky:1985xj}. 
Wilson line correlators are then governed by renormalisation group equations,
involving an anomalous dimension. For the case of a single cusp, the latter is
known as the {\it cusp anomalous dimension}, which has become a quantity of
central importance in the study of infrared singularities and resummation.\\

In Abelian theories, the all-order structure of the cusp anomalous dimension
is dictated by a simple subset of possible Feynman diagrams, namely those in
which the Wilson line contours meeting at the cusp are joined by a connected
subgraph. The only possibilities are single photon emissions, or photon 
emissions which are joined by fermion loops. Traditionally this result was 
derived using the eikonal identity for multiple soft photon 
emissions~\cite{Yennie:1961ad}. Recently, an alternative proof has been given,
using path integral techniques that relate the exponentiation of connected
subgraphs in Wilson loops with a cusp, to the textbook 
exponentiation of connected diagrams in a general quantum field 
theory~\cite{Laenen:2008gt}. This simple structure is referred to as 
{\it Abelian exponentiation} in the literature, and in particular implies that
if there are no propagating fermions, the cusp anomalous dimension is one-loop
exact. Consequently, the infrared singularities of scattering amplitudes are
purely governed by the exponentiation of the one-loop result. A similar result
holds for correlators involving more than two Wilson lines meeting at a point.
Then, the relevant anomalous dimension is completely determined to all orders
by connected subdiagrams spanning the outgoing particles. Again, one-loop
exactness follows if fermion loops are absent.\\

In non-Abelian theories, the cusp anomalous dimension is, unsurprisingly, more
complicated. Nevertheless, its all order structure is still dictated by a 
subset of Feynman diagrams, which permit a simple topological classification:
as well as connected subdiagrams spanning the outgoing Wilson line contours,
one may also have diagrams which are {\it two-eikonal line 
irreducible}\footnote{That is, subdiagrams which cannot be disconnected by
a single cut through both eikonal lines.}. Such
diagrams have modified colour factors which are maximally non-Abelian, and are
referred to as {\it webs} in the 
literature~\cite{Gatheral:1983cz,Frenkel:1984pz,Sterman:1981jc}\footnote{A 
novel geometric interpretation of webs in two parton scattering has very 
recently been presented in~\cite{Erdogan:2011yc}.}. Another way
to think about webs is that they are, by definition, those diagrams which enter
the exponent of the Wilson line correlator. Their structure is thus usually 
referred to as {\it non-Abelian exponentiation}, by analogy with the Abelian 
case. These results were recently reconsidered in~\cite{Laenen:2008gt}, where 
an alternative derivation of webs was given using statistical physics methods. 
Furthermore, the notion of webs has recently been extended to correlators 
involving many Wilson lines meeting at a point, by two groups of 
authors~\cite{Mitov:2010rp,Gardi:2010rn}. By analogy with the two line case
(alternatively, a single contour with a cusp), webs are those diagrams which
enter the exponent of the Wilson line correlator. However, their structure is
markedly different to the two line case. The most significant difference is 
that {\it reducible} diagrams contribute to the exponent in general. 
Furthermore, webs become closed sets of diagrams whose members are related by
gluon permutations. Their colour and kinematic information is entangled to all
orders in perturbation theory by so-called {\it web mixing 
matrices}~\cite{Gardi:2010rn}, whose structure is in principle governed purely
from combinatorics~\cite{Gardi:2011wa}. This structure has yet to be fully 
explored, but is certainly crucial to understanding the all-order structure
of infrared singularities in multileg scattering 
amplitudes~\cite{Gardi:2011yz}.\\

Recently, a new way to think about Wilson lines was presented 
in~\cite{Chien:2011wz}. The authors start off by considering a collection of
Wilson lines meeting at a point in Minkowski space. They then transform to a 
set of {\it radial coordinates}, which map Minkowski space to
$\mathbb{R}\times{\rm AdS}$, where ${\rm AdS}$ denotes Euclidean 
anti-de-Sitter space. Parametrisation invariance of the Wilson line contours
in Minkowski space becomes time translation invariance in the transformed 
space, such that each Wilson line can be considered as a static charge. In 
addition, the dilatation operator in Minkowski space (whose eigenvalues are
dimensions) maps to the operator $\partial_\tau$ in radial coordinates, 
where $\tau$ is the time-like coordinate normal to the (space-like) slices of 
AdS. Eigenvalues of the latter operator are energies (up to a factor of $i$),
so that one may obtain the anomalous dimension associated with a given Wilson 
line correlator by evaluating the energy of a collection of static charges in
Euclidean AdS space. This is worked out in detail in~\cite{Chien:2011wz} for 
QED and QCD at one loop, and the conceptual picture thus obtained is used to 
motivate a family of {\it conformal gauges} for use in field theory Wilson line
calculations. At two loops, as shown explicitly in~\cite{Chien:2011wz}, such 
gauges eliminate non-Abelian graphs containing a three-gluon vertex. Although 
such graphs are not completely absent at higher orders, it seems likely that 
such gauges may provide a natural framework for probing the conjectured 
all-order structure of IR singularities in QCD fixed-angle scattering 
amplitudes~\cite{Gardi:2009qi,Becher:2009cu,Becher:2009qa}, which involves 
correlations between only pairs of partons in the case that they are all 
massless. \\

The purpose of this paper is to extend the radial picture for Wilson lines
to perturbative quantum gravity, in the form of General Relativity (GR) 
minimally coupled to (scalar) matter. The consideration of infrared 
singularities in gravity dates back to~\cite{Weinberg:1965nx}, and there has 
recently been renewed interest~\cite{Naculich:2011ry,White:2011yy,
Akhoury:2011kq}, based on a number of motivations. Firstly, although GR 
contains non-renormalisable ultraviolet singularities, alternative field 
theories of gravity may have the same long distance 
behaviour~\cite{Giddings:2010pp}. Secondly, there is a growing body of work
involving intriguing connections between scattering amplitudes in gauge and 
gravity theories, at arbitrary loop level (see e.g.~\cite{Bern:2010yg,
Bern:2010ue,Bern:2002kj}, and~\cite{BoucherVeronneau:2011qv,Naculich:2011my} 
for recent applications), which may have a string theoretic origin. There 
then exists the real possibility that knowledge about quarks and gluons can 
tell us about gravity, or vice versa. To this end, it is important to develop 
common conceptual pictures, that allow us interpret similar physics (such as 
infrared singularity structures) in a natural way, an observation which 
motivates the present study.\\

The structure of the paper is as follows. In section~\ref{sec:review} we review
the radial picture for Wilson lines developed in~\cite{Chien:2011wz}, as 
well as recent results from perturbative quantum gravity, which will be 
necessary for the rest of the paper. In section~\ref{sec:wilson}, we apply the
radial picture to Wilson line operators in quantum gravity, showing 
explicitly that the form of the one-loop cusp anomalous dimension in gravity
can be reproduced from the Newtonian limit of General Relativity 
in Euclidean AdS space. In section~\ref{sec:collinear}, we comment on the 
lightlike limit of gravitational Wilson line operators, in particular 
observing the cancellation of collinear singularities in the radial picture. 
In section~\ref{sec:gen} we introduce a general formulation of the cusp
anomalous dimension calculation, and show that a continuous parameter exists
that interpolates smoothly between the QED and gravity cases.
In section~\ref{sec:conclusion}, we discuss our results before concluding.
Some technical details are collected in appendices. 

\section{Review of necessary concepts}
\label{sec:review}
\subsection{The radial picture for Wilson lines}
In this section, we briefly review the approach of~\cite{Chien:2011wz} for 
describing Wilson lines as static charges in AdS space. Given that we will
be focussing explicitly on gravity in the rest of the paper, we here discuss
only QED, thus ignoring additional complications due to non-trivial colour
structure. \\

Let us start with the definition of a Wilson line operator in an Abelian gauge
theory:
\begin{equation}
{\cal W}({\cal C})=\exp\left[ie\int_{\cal C}dx_\mu A^\mu(x) \right],
\label{Wilson}
\end{equation}
where $A^\mu(x)$ is the gauge field, $e$ the coupling constant, and the line 
integral is taken over the contour ${\cal C}$ in Minkowski space. In 
applications of Wilson lines to scattering amplitudes, one is interested in
Wilson line contours which are fully determined by the momenta of the outgoing
particles. Each Wilson line then represents an outgoing particle dressed by 
an infinite number of soft photon emissions, and the soft part of the 
scattering amplitude is given by the vacuum expectation value of the product
\begin{equation}
{\cal W}(n_1,\ldots,n_L)=\prod_{i=1}^L\exp\left[ie\int_0^\infty ds_i p_i\cdot 
A(sp_i) \right],
\label{Wilsonprod}
\end{equation}
where $L$ is the number of hard outgoing particles. 
Here we have parameterised each (straight) Wilson line contour by $x_i=s_ip_i$,
where $p_i$ is the 4-momentum of outgoing particle $i$. Note that $s_i$ then 
has dimensions of (length)$^2$. It is more common in the literature to instead
write $x_i=t_in_i$, where $n_i$ is the 4-velocity. The parameter $t_i$
then has dimensions of length, but one is free to make the above choice, given
that the Wilson line operator is invariant under rescalings
\begin{equation}
p_i\rightarrow\frac{p_i}{\lambda},\quad s_i\rightarrow\lambda s_i.
\label{rescale}
\end{equation}
The fact that we choose to use the momentum rather than the velocity in 
eq.~(\ref{Wilson}) is to make explicit the analogy between gauge theory and
gravity, where in the latter theory momenta are more important given that they
play the role of charges. \\

Consider now Minkowski space with time coordinate $t$, and spherical polar
coordinates $r$, $\theta$ and $\phi$, representing the radial, polar and 
azimuthal coordinates respectively. Then a given Wilson line direction can be
parametrised by~\cite{Chien:2011wz}
\begin{equation}
x^\mu=e^\tau(\cosh\beta,\sinh\beta\hat{\vec{n}}),
\label{Wildir}
\end{equation}
where $\hat{n}$ is a unit 3-vector. 
In these so-called {\it radial coordinates}, the Minkowski space metric 
becomes\footnote{Note that in this subsection only we use the metric 
(+,--,--,--), for
ease of comparison with~\cite{Chien:2011wz}. Throughout the rest of the paper,
we will use the alternative choice (--,+,+,+), which is more common in the
gravity literature.}
\begin{equation}
ds^2=e^{2\tau}\left[d\tau^2-(d\beta^2+\sinh^2\beta\, d\Omega_2^2)\right],
\label{metric}
\end{equation}
where $d\Omega_2^2$ is the squared line element on a 2-sphere. Next, one 
interprets $-\infty\leq\tau\leq\infty$ as a time coordinate, and uses the 
fact that Abelian gauge theory (in the absence of propagating fermions) is 
classically conformally invariant in four dimensions. That is, one is free to 
rescale the line element of eq.~(\ref{metric}) by an overall factor, and thus 
to replace eq.~(\ref{metric}) by
\begin{equation}
ds^2=d\tau^2-(d\beta^2+\sinh^2\beta\, d\Omega_2^2).
\label{metric2}
\end{equation}
The time coordinate is now explicitly decoupled from the spatial coordinates,
and spatial slices (i.e.\ at fixed $\tau$) constitute Euclidean AdS space in 
three dimensions. Two important identifications between the original Minkowski
space and the transformed space are as follows:
\begin{itemize}
\item Reparametrisations of the Wilson line contour in Minkowski space, e.g.\
the rescalings of eq.~(\ref{rescale}), map to time translations 
in the transformed space. 
Thus, reparametrisation invariance of Wilson lines shows up, in the latter
space, as invariance under shifts in $\tau$. This allows one to consider the
Wilson lines in the transformed space as static charges.
\item The dilatation operator $x^\mu\partial_\mu$ (whose eigenvalues are 
dimensions) maps to the operator $\partial_\tau$ in the transformed space.
This is related to the Hamiltonian operator in the transformed space via
\begin{equation}
\partial_\tau=i{\cal H}^{\mathbb{R}\times{\rm AdS}},
\label{Htau}
\end{equation}
so that the anomalous dimension of a Wilson line correlator maps to the
energy of the corresponding collection of static charges in the Euclidean AdS
space, with an additional factor of $i$. 
\end{itemize}
The electrostatic potential $\phi$ due to the two charges is given by the 
solution of Laplace's equation in Euclidean AdS space:
\begin{equation}
\nabla^2\phi=\frac{1}{\sinh^2\beta}\partial_\beta\left(\sinh^2\beta(
\partial_\beta\phi)\right)=0,
\label{laplace}
\end{equation}
whose solution is
\begin{equation}
\phi(\beta)=C_1+C_2\coth\beta,
\label{lapsol}
\end{equation}
where the $\{C_i\}$ are constants of integration. As explained in detail 
in~\cite{Chien:2011wz}, this does not have the correct behaviour as 
$\beta\rightarrow\infty$, where the cusp anomalous dimension should diverge
linearly in $\beta$, due to the appearance of collinear singularities in 
Minkowski space. This is due to the fact that the solution implicitly includes
the effect of a spurious charge corresponding to a phantom initial state 
particle (i.e. eq.~(\ref{lapsol}) has a pole at $\beta=0$ and $\beta=i\pi$).
One may correct for this by first analytically continuing the 
spacelike part of the radial coordinate space to a Euclidean 3-sphere, and 
subsequently adding a constant charge density to the source term of the 
equation for the electrostatic potential. These constant densities cancel out
upon constructing any collection of Wilson line operators, which must satisfy
charge conservation. The corrected solution is
\begin{equation}
\phi=\frac{1}{4\pi^2}\left[(\pi+i\beta)\coth\beta+C\right],
\label{lapsol2}
\end{equation}
where the overall normalisation is fixed by the amount of constant charge 
necessary to cancel the phantom charges. \\

Having constructed the potential, the total energy of a pair of charges (which
in~\cite{Chien:2011wz} is obtained by first constructing the electric field) 
is 
\begin{equation}
E(\beta_{12})=\frac{q_1q_2}{4\pi^2}\left[(\pi+i\beta_{12})\coth\beta_{12}
+C\right],
\label{Etot}
\end{equation}
where $\beta_{12}$ is the geodesic distance between the charges, which is 
identified with the cusp angle in Minkowski space. The constant $C$ can be 
fixed by analytic continuation to the situation in which one of the Wilson 
lines is incoming. The vanishing of the cusp anomalous dimension when the two
Wilson lines become parallel then implies that $C=-i$~\cite{Chien:2011wz}. 
Finally, one finds that the cusp anomalous dimension should be given by
\begin{equation}
\Gamma=iE(\beta_{12})=\frac{q_1q_2}{4\pi^2}\left[(i\pi-\beta_{12})
\coth\beta_{12}+1\right],
\label{cusp1}
\end{equation}
in exact agreement with an explicit Minkowski space field theory calculation.
The above argument is for the case of QED. The generalisation to a non-Abelian
context is straightforward, and is also presented in~\cite{Chien:2011wz}.

\subsection{Infrared singularities in Quantum GR}
In the previous section, we have reviewed the arguments of~\cite{Chien:2011wz}
which relate the cusp anomalous dimension in a gauge theory to the energy of
a static charge configuration in Euclidean AdS space. The aim of the present
paper is to outline how the same arguments can be applied in quantum gravity.
To this end, we must first recap salient features regarding the structure of
infrared singularities in GR. For ease of comparison with previous literature 
on quantum gravity, we will from now on adopt the (--,+,+,+) metric, in 
contrast with the previous subsection.\\

Here we will consider general relativity minimally coupled to a scalar field
$\Phi$, a theory whose total action is given by
\begin{equation}
S=S_{\rm E.H.}[g^{\mu\nu}]+S_{\rm mat}[\Phi^*,\Phi,g^{\mu\nu}],
\label{Stot}
\end{equation}
where $S_{\rm E.H.}$ is the Einstein-Hilbert action, and 
\begin{equation}
S_{\rm mat}[\Phi^*,\Phi,g^{\mu\nu}]=\int d^dx\sqrt{-g}\left[-g^{\mu\nu}
\partial_\mu\Phi^*\partial_\nu\Phi-m^2\Phi^*\Phi\right]
\label{Smat}
\end{equation}
in $d$ dimensions, where $g$ is the determinant of the metric tensor 
$g_{\mu\nu}$. Perturbation theory can be defined by expanding the metric tensor
according to
\begin{equation}
g_{\mu\nu}=\eta_{\mu\nu}+\kappa h_{\mu\nu},
\label{gmunuexp}
\end{equation}
where $h_{\mu\nu}$ is then the graviton field, and $\kappa=\sqrt{16\pi G_N}$, 
with $G_N$ Newton's constant. Note that a different choice is often made in 
the literature (see e.g.~\cite{Capper:1973bk}), in which one instead defines 
the graviton via the expansion of the quantity
\begin{equation}
\tilde{g}^{\mu\nu}=\sqrt{-g}g^{\mu\nu}.
\label{gtilde}
\end{equation}
This simplifies the Feynman rules for emission of gravitons from scalar 
particles, and was the choice adopted by~\cite{White:2011yy} which derived
Wilson line operators for soft graviton emission. Here, however, we will stick
with the choice of eq.~(\ref{gmunuexp}), for reasons that will become clear.
Substituting eq.~(\ref{gmunuexp}) into eq.~(\ref{Smat}) and expanding up to
first order in $\kappa$ (including also the factor $\sqrt{-g}$), 
the matter action becomes\footnote{Note that after 
expanding in the weak field approximation, contractions involving upper and
lower indices are interpreted as involving the Minkowski metric 
$\eta_{\mu\nu}$.}
\begin{align}
S_{\rm mat}[\Phi^*,\Phi,h^{\mu\nu}]&=\int d^dx\left[-\partial^\mu\Phi^*
\partial_\mu\Phi-m^2\Phi^*\Phi+\kappa\left(-\frac{m^2}{2}\eta^{\mu\nu}
\Phi^*\Phi
-\frac{1}{2}\eta^{\mu\nu}\partial_\alpha\Phi^*\partial^\alpha\Phi
\right.\right.\notag\\
&\left.\left.\phantom{\frac{1}{2}}-\partial^\mu\Phi^*\partial^\nu\Phi\right)h_{\mu\nu}\right]+{\cal O}(\kappa^2).
\label{Smat2}
\end{align}
The momentum-space Feynman rule for single graviton emission from a scalar is
then
\begin{equation}
\frac{i\kappa}{2}\left[(-m^2+p_1\cdot p_2)\eta^{\mu\nu}+p_1^\mu\,p_2^\nu
+p_1^\nu\,p_2^\mu\right],
\label{FR1g}
\end{equation}
where $p_1$ and $p_2$ are both outgoing. We will also need the graviton 
propagator, which in $d$ dimensions is given (in the de Donder gauge) 
by~\cite{Veltman:1975vx}
\begin{equation}
D_{\mu\nu,\alpha\beta}(k)=\frac{-iP_{\mu\nu,\alpha\beta}}{k^2-i\epsilon},\quad
P_{\mu\nu,\alpha\beta}=\eta_{\mu\alpha}\,\eta_{\nu\beta}
+\eta_{\mu\beta}\,\eta_{\nu\alpha}-\frac{2}{d-2}\eta_{\mu\nu}\,
\eta_{\alpha\beta}. 
\label{prop}
\end{equation}

Given the theory defined by eq.~(\ref{Stot}), one may consider scattering 
amplitudes involving a number $L$ of external scalars, which interact by the 
exchange of gravitons. When emitted gravitons become soft, infrared 
divergences occur. The study of infrared singularities in quantum general 
relativity dates back to the classic paper~\cite{Weinberg:1965nx}, which 
established that Abelian exponentiation can be generalised to gravity from 
the QED case. Earlier this year, attention again turned to the IR sector of 
gravity, with a particular emphasis on describing the singularity structure 
using methods developed in the context of gauge theory~\cite{Naculich:2011ry,
White:2011yy,Akhoury:2011kq}, leading both to an extension of our knowledge of
IR effects in gravity, and also a more unified view of the shared physics 
underlying the IR sectors of gravity and gauge theories. It has been known for
some years that amplitudes in gauge theory with $L$ external lines factorise 
into the schematic form~\cite{Mueller:1979ih,Collins:1980ih,Sen:1981sd,
Korchemsky:1988pn,Magnea:1990zb}
\begin{equation}
{\cal A}_L={\cal S}\cdot{\cal H}\cdot\prod_{i=1}^L J_i.
\label{ALdef}
\end{equation}
Here ${\cal S}$ is a {\it soft function}, collecting all infrared singularities
due to soft gauge boson emission, ${\cal H}$ is a {\it hard function} which is
finite in dimensional regularisation as $d\rightarrow4$, and $J_i$ is a
{\it jet function} collecting hard collinear singularities associated with
the $i^{\rm th}$ external line\footnote{Strictly speaking, there is a double counting in
eq.~(\ref{ALdef}), given that soft-collinear singularities are present in both
the jet functions and the soft function. This can be easily rectified by 
dividing by {\it eikonal jet functions} ${\cal J}_i$, which we may ignore in
the present discussion.}. The soft function is given by a vacuum expectation
value of Wilson lines meeting at a point, where there is one Wilson line for
each external particle. The authors of~\cite{Naculich:2011ry} suggested that
a similar structure should hold for pure quantum gravity, namely that one has
\begin{equation}
{\cal A}_L={\cal S}\cdot{\cal H}
\label{ALdef2}
\end{equation}
in the gravitational context, where again ${\cal S}$ and ${\cal H}$ denote 
soft and hard functions, but in this case there are no jet functions, due to 
the fact that collinear singularities cancel after summing over diagrams.
The latter fact was first established in~\cite{Weinberg:1965nx}, for 
soft-collinear emissions. As in the gauge theory case, the soft function is
given by a Wilson line correlator, where an appropriate form of the Wilson
line operator (representing the emission of soft gravitons from an eikonal 
line) was suggested in~\cite{Naculich:2011ry}\footnote{Note that what we 
refer to as a Wilson line in this paper is not to be confused with the 
parallel transport operator of GR, defined in terms of the Christoffel symbol,
which is also sometimes referred to in Wilson line terms.}. Furthermore, the
anomalous dimension governing the renormalization of gravitational Wilson line
correlators was argued to be one loop exact. Reference~\cite{White:2011yy} 
investigated some of these issues further in the context of GR 
minimally coupled to matter (as described by eq.~(\ref{Stot})), and derived 
explicit forms for the Wilson line operators of quantum gravity using path 
integral techniques borrowed from gauge theory~\cite{Laenen:2008gt}. This 
also generalised the discussion to include the effects of massive eikonal 
lines, and corrections beyond the eikonal approximation. Finally, the authors
of~\cite{Akhoury:2011kq} have undertaken a complete and rigorous analysis of 
the infrared singularity structure of pure gravity amplitudes, using techniques
earlier applied to QCD (see~\cite{Sterman:1995fz} for a pedagogical review).
This confirmed the one-loop exactness of the anomalous dimension underlying
the IR singularity structure of gravity amplitudes. In addition, the 
cancellation of collinear singularities was generalised from soft to hard
collinear configurations.\\

\begin{figure}
\begin{center}
\scalebox{1.0}{\includegraphics{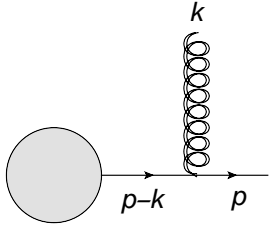}}
\caption{The emission of a graviton from an external scalar line.}
\label{fig:1gv}
\end{center}
\end{figure}

For this paper, we will need the form of the gravitational Wilson line 
operator for the weak field expansion of eq.~(\ref{gmunuexp}), for a straight
line contour along the direction of an outgoing particle. This is given
by~\footnote{An alternative Wilson line operator was studied 
in~\cite{Brandhuber:2008tf}. See also~\cite{Naculich:2011ry} for a discussion 
of this point.}
\begin{equation}
{\cal W}_g=\exp\left[\frac{i\kappa}{2}\int_0^\infty ds \,p^\mu p^\nu\,
h_{\mu\nu}(sp^\mu)\right],
\label{Wgdef}
\end{equation}
for momentum $p^{\mu}$. Note that this is not the same as the result presented
in~\cite{White:2011yy}, owing to the fact that that paper used a weak field
expansion based on eq.~(\ref{gtilde}). However, we can justify this result as
follows. First, note that the emission of a graviton from an external scalar
line is given by figure~\ref{fig:1gv}. 
This introduces a propagator for the intermediate scalar line, and an emission
vertex for the graviton. The latter is given by the vertex of eq.~(\ref{FR1g})
with $p_1=-(p-k)$ and $p_2=p$ (recalling that the $\{p_i\}$ were defined to be
outgoing above), which in the soft limit $k\rightarrow0$ becomes
\begin{equation}
\frac{i\kappa}{2}\left[-(p^2+m^2)\eta^{\mu\nu}-2p^\mu\,p^\nu\right]=
-i\kappa\,p^\mu\,p^\nu.
\label{eikrule1}
\end{equation}
Combining this with the propagator for the intermediate scalar line gives
\begin{equation}
\frac{-i}{(p-k)^2+m^2-i\epsilon}(-i\kappa\,p^\mu\,p^\nu)=\frac{\kappa}{2}
\frac{p^\mu\,p^\nu}{p\cdot k}.
\label{eikrule2}
\end{equation}
The right-hand side constitutes an effective Feynman rule for the emission of
a soft graviton from an eikonal line. In particular, it is insensitive to the
spin of the emitting particle. This same Feynman rule is indeed generated by 
the Wilson line operator of eq.~(\ref{Wgdef}), which shows that the latter is
the correct operator.  \\

An interesting consistency check of the above Wilson line operator is that
one may derive Newton's law of gravity, by calculating the expectation value
of a certain Wilson loop in Minkowski space. This is analogous to the 
derivation of the Coulomb potential in QED. Although the derivation of Newton's
law from perturbative quantum gravity is by no means a new 
result~\cite{Veltman:1975vx}, this calculation is usually performed by taking
the non-relativistic limit of the full one-graviton exchange graph. Given that
the alternative derivation gives extra insight into the gravitational Wilson
line operators, we present this here in appendix~\ref{app:newton}\footnote{This
calculation is closely related to a similar analysis in~\cite{Hamber:1994jh}, 
as we explain in the appendix.}.\\

Armed with the above results, we are now ready to examine Wilson line operators
in quantum gravity, using the Euclidean AdS picture of~\cite{Chien:2011wz}. 
This is the subject of the following section.

\section{Gravitational Wilson lines in the radial picture}
\label{sec:wilson}
In the previous sections, we have reviewed both the radial coordinate picture
for calculating cusp anomalous dimensions of Wilson line operators, and also
the known structure of infrared divergences in quantum GR. In this section,
we combine these concepts, in order to elucidate the gravitational analogue
of the analysis of~\cite{Chien:2011wz} for abelian and non-abelian gauge 
theories. As stated in the previous section, we use the metric (--,+,+,+), so 
that the equivalent of the radial coordinate space of eq.~(\ref{metric}) is
\begin{equation}
ds^2=e^{2\tau}[-d\tau^2+(d\beta^2+\sinh^2\beta\, d\Omega_2^2)].
\label{metric-}
\end{equation}

Our starting point is to note that, unlike the case of conventional gauge 
theories, the Wilson line operator of eq.~(\ref{Wgdef}) is not invariant 
under rescalings of the form of eq.~(\ref{rescale}). This does not prevent us
from examining Wilson lines in radial coordinates. However, it is not then 
true that the system in the transformed space will be invariant under time
translations (in $\tau$). In particular, one must include the prefactor of 
$e^{2\tau}$ in the metric of eq.~(\ref{metric-}), which changes the nature
of computations involving energies in radial space. In calculating
the latter, we will be interested in the component $h_{\tau\tau}$ of the 
graviton field. The lack of rescaling invariance in Minkowski space implies 
that this will be $\tau$-dependent. We can surmise this dependence by 
rewriting the Wilson line operator of eq.~(\ref{Wgdef}) as
\begin{equation}
{\cal W}_g=\exp\left[\frac{i\kappa}{2}p^\mu\int_0^\infty ds \,p^\nu\,
h_{\mu\nu}(sp^\mu)\right],
\label{Wgdef2}
\end{equation}
where symmetry of the graviton field ($h_{\mu\nu}=h_{\nu\mu}$) means that we
can decide to take either of the momentum factors out of the integral along the
Wilson line parameter. The exponent now has the explicit form of a 
reparametrisation-invariant term
\begin{equation}
\int dx^\nu\,h_{\mu\nu}(x^\mu)
\label{inv1}
\end{equation}
multiplied by a charge factor $p^\mu$. That this charge is the 4-momentum of
the emitting particle is as expected from general relativity. Rewriting 
the expression~(\ref{inv1}) as 
\begin{equation}
\int_0^\infty d\tau \,\dot{x}^\nu\,h_{\mu\nu}(x^\mu),
\label{inv2}
\end{equation}
where the dot represents differentiation with respect to $\tau$, it follows 
that 
\begin{equation}
\frac{\partial}{\partial \tau}\dot{x}^\nu h_{\mu\nu}=
\frac{\partial}{\partial \tau}h_{\mu\tau}=0,
\label{inv3}
\end{equation}
so that the invariance under $\tau$ translations results from the invariance
of the expression~(\ref{inv2}) under reparametrisations. 
That is, the quantity $h_{\tau\nu}$ is independent of $\tau$ in the radial
coordinate space~\footnote{Care is needed here in interpreting the notation.
The index $\nu$ in $h_{\tau\nu}$ is a Minkowski space index, which has yet
to be transformed to the radial coordinate space.}. As a consequence, one may 
write
\begin{equation}
h_{\tau\tau}=\dot{x}^\mu \dot{x}^\nu h_{\mu\nu}=e^\tau\,K(\beta),
\label{httform}
\end{equation} 
where we have used the fact that $\dot{x}^\mu=x^\mu=e^\tau n^\mu$, with $n^\mu$
independent of $\tau$. Given that $n^\mu$ is in the direction of the hard 
momentum $p^\mu$ of the particle emitting soft gravitons, we may write
\begin{equation}
n^\mu=\frac{p^\mu}{m},\quad x^\mu=e^\tau\frac{p^\mu}{m},
\label{xmudef}
\end{equation}
where the normalisation ensures $n^2=-1$. \\

Analogously to the QED analysis of~\cite{Chien:2011wz}, one may consider 
calculating the potential energy of a pair of charges in the radial coordinate
space. In this case, these will be masses, which we label by $m_1$ and
$m_2$. The total potential energy can be constructed by first considering the
potential energy generated by the mass $m_2$ located at the origin of the 
radial coordinate space, and then considering the mass $m_1$ as a test particle
a geodesic distance $\beta$ away.\\

To this end, we must construct the equation satisfied by $h_{\tau\tau}$ in 
radial coordinate space, and the first step is to note that the Wilson line 
phase of eq.~(\ref{Wgdef2}) can be written as
\begin{equation}
\frac{i\kappa}{2}p^\nu\int_0^\infty dx^\mu \,
h_{\mu\nu}(sp^\mu)=\frac{i\kappa}{2}p^\nu\int_{-\infty}^{\infty}d\tau
\,h_{\tau\nu}.
\label{jmunu1}
\end{equation}
Inserting a delta-function in the radial coordinate space, this can be further
rewritten as
\begin{equation}
\frac{i\kappa}{2}p^\nu\int_{-\infty}^{\infty}d\tau\int d^3\vec{x}
\,\delta^{(3)}(\vec{x})\,h_{\tau\nu}.
\label{jmunu2}
\end{equation}
Equating this to the general form of a source term in curved space,
\begin{displaymath}
i\int d^4x\sqrt{-g}j^{\mu\nu}h_{\mu\nu},
\end{displaymath}
one finds that the current which sources the conformally invariant quantity
$h_{\tau\nu}$ is given by
\begin{equation}
\sqrt{-g}j^{\tau\nu}=\frac{\kappa}{2}\,p^\nu\,\delta^{(3)}(\vec{x}).
\end{equation}
One then finds
\begin{equation}
j_{\mu\tau}=g_{\tau\tau}j_\mu^\tau=-\frac{\kappa}{2}e^{-2\tau}p_\mu
\delta^{(3)}(\vec{x}),
\label{jmunu4}
\end{equation}
where we have absorbed geometric factors into the delta function so that this
is now normalised according to
\begin{equation}
\int d^3\vec{x}\sqrt{g^{(3)}}\delta^{(3)}(\vec{x})=1, 
\label{deltnorm}
\end{equation}
where $g^{(3)}$ is the determinant of the spatial part of the metric 
$g_{\mu\nu}$, disregarding the overall factor of $e^{2\tau}$.
The component $h_{\tau\tau}$ of the graviton field satisfies the wave equation
(see e.g.~\cite{Landau})
\begin{equation}
\Box h_{\tau\tau}=j_{\tau\tau},
\label{htteq}
\end{equation}
where from eqs.~(\ref{xmudef}, \ref{jmunu4}) one finds
\begin{equation}
j_{\tau\tau}=\dot{x}^\mu\,j_{\mu\tau}=\frac{\kappa}{2}m_2e^{-\tau}\delta^{(3)}
(\vec{x}).
\end{equation}
The left-hand side of eq.~(\ref{htteq}) is the covariant D'Alambertian operator
\begin{equation}
\Box=\frac{1}{\sqrt{-g}}\partial_\mu\left(\sqrt{-g}g^{\mu\nu}\partial_\nu
\right).
\label{dalambert}
\end{equation}
We see from eq.~(\ref{htteq}) that gravitational Wilson lines in the 
radial space correspond to time-dependent sources. This is a manifestation of 
the overall $e^{2\tau}$ factor in the metric of eq.~(\ref{metric-}): 
the radial size of the spacelike part of the space changes with $\tau$. If we 
consider a mass fixed at the origin, the flux of the gravitational field 
through an expanding surface of radius $\sim e^{\tau}\beta$ will be constant. 
If we instead consider a surface at fixed $\beta$ around the origin, the flux 
through such a surface grows as a function of time, such that it looks as if 
a mass $me^{-\tau}$ is enclosed\footnote{One must also remember that there is 
a factor $\sqrt{-g}$ involved in calculating the flux, so that a mass which 
decreases with time is consistent with a flux which grows with time.}, which 
is precisely the content of eq.~(\ref{htteq}). \\

Using the metric of eq.~(\ref{metric-}), eq.~(\ref{htteq}) becomes
\begin{equation}
e^{-2\tau}\left(\nabla^2-3\right)h_{\tau\tau}=
\frac{\kappa}{2}m_2e^{-\tau}\delta^{(3)}(\vec{x}),
\label{htteq2}
\end{equation}
where $\nabla^2$ is the three-dimensional Laplacian\footnote{Note that, for 
ease of comparison with the QED case, we denote by $\nabla^2$ the Laplacian
without the $\tau$ dependence implied by the metric of eq.~(\ref{metric-}).}. 
The homogeneous equation is not simply Laplace's equation as in the case of 
QED, but rather the Helmholtz equation. The additional term linear in 
$h_{\tau\tau}$ is a direct consequence of the fact that the flux through a 
surface of fixed $\beta$ is changing with time due to the overall time 
dependence of the metric, as discussed above. The relation of the graviton 
field to the Newtonian potential is (again see e.g.~\cite{Landau})
\begin{equation}
\Phi=\frac{\kappa}{2}h_{\tau\tau},
\label{newtonphi}
\end{equation}
so that eq.~(\ref{htteq2}) may be rewritten as
\begin{equation}
\nabla^2\Phi-3\Phi=
\left(\frac{\kappa}{2}\right)^2m_2e^{\tau}\delta^{(3)}(\vec{x}).
\label{htteq3}
\end{equation}
This is the gravitational analogue of Laplace's equation for the QED case 
of~\cite{Chien:2011wz}. It is essentially Newton's law of gravity, but with an
extra term proportional to $\Phi$ resulting from the use of a time-dependent
curvilinear coordinate system. From eqs.~(\ref{httform}, \ref{newtonphi}) we 
may write
\begin{equation}
\Phi(\beta,t)=\tilde{K}(\beta)e^\tau,\quad\tilde{K}(\beta)=\frac{\kappa}{2}
K(\beta).
\label{Ktildedef}
\end{equation}
Then using the Laplacian of eq.~(\ref{laplace}), we must find the general 
solution of the homogeneous equation 
\begin{equation}
\frac{1}{\sinh^2\beta}\partial_\beta\left(\sinh^2\beta\partial_\beta
\tilde{K}(\beta)\right)-3\tilde{K}(\beta)=0,
\label{Ktildeeq}
\end{equation}
which is
\begin{equation}
\tilde{K}(\beta)=A_1\left[\frac{1}{\sinh\beta}+2\sinh\beta\right]
+A_2\cosh\beta.
\label{gensolK}
\end{equation}
Note that, as in the QED case, the general solution of the homogeneous equation
is a superposition of a function which is even under the transformation of
eq.~(\ref{betatrans}), and a function which is odd.
These are the first and second terms in eq.~(\ref{gensolK}) respectively.
As shown in appendix~\ref{app:norm}, one can fix $A_1$ by considering a 
3-surface consisting of a cylinder of radius $\beta_0$ and height (in time) 
$\tau_0$. The result is
\begin{equation}
\tilde{K}(\beta)=m_2\frac{\kappa^2}{16\pi}\left[\frac{1}{\sinh\beta}
+2\sinh\beta\right]+A_2\cosh\beta.
\label{gensolK2}
\end{equation}
Another feature shared with the QED case is that this solution has the wrong
behaviour as $\beta\rightarrow\infty$. As shown in appendix~\ref{app:cuspcalc},
the actual result $\sim \beta e^\beta$ at large $\beta$. This is a factor 
$e^\beta$ up on the QED case (where the divergence is linear), due to the fact
that there are extra momentum factors in gravity. The failure of the 
solution~(\ref{gensolK2}) to reproduce this behaviour is, as in the QED case, 
due to a spurious charge. Noting that eq.~(\ref{gensolK2}) has a pole
at both $\beta=0$ and $\beta=i\pi$, this solution represents the combined 
effect of the physical mass at $\beta=0$, and a spurious mass at 
$\beta=i\pi$.\\ 

In~\cite{Chien:2011wz}, and as reviewed in section~\ref{sec:review}, this 
problem is solved in the QED case by modifying the current $j_\tau$ by a 
constant charge density. A similar procedure can be used in the gravity case,
as we now discuss. First, recall that the current $j_{\mu\tau}$ is given by
eq.~(\ref{jmunu4}), and that this sources the {\it conformally invariant} 
quantity $h_{\mu\tau}$ involving one Minkowski-space and one radial-space 
index. If we were to solve for
$h_{\mu\tau}$, we would find a charge $p^\mu$ at the origin $\beta=0$ 
of the radial space, and a spurious charge $-p^\mu$ (corresponding to an 
incoming momentum in Minkowski space) at $\beta=i\pi$. The analogue of adding a
constant charge density in the present case is to modify the current 
of eq.~(\ref{jmunu4}) so as to give
\begin{equation}
j_{\mu\tau}=-\frac{\kappa}{2}e^{-2\tau}\,p_\mu\left[\delta^{(3)}(\vec{x})
+K\right],
\label{jmutau2}
\end{equation}
where $K$ is such as to remove the spurious incoming momentum. Any given 
collection of gravitational Wilson lines will obey momentum conservation,
such that the constant charge densities thus added will cancel out. This is
the analogue of the cancellation of the constant terms due to electric charge
conservation in the QED case. \\

Above, we solved directly for $h_{\tau\tau}$, which is related to the Newtonian
potential in the radial coordinate space. To this end, one must consider the
modified current
\begin{equation}
j_{\tau\tau}=\dot{x}^\mu j_{\mu\tau}=\frac{\kappa}{2}e^{-2\tau}
\dot{x}^\mu\,p_\mu \left[\delta^{(3)}(\vec{x})+K\right].
\label{jtautau2}
\end{equation}
Here $\dot{x}^\mu=x^\mu=e^{\tau}(\cosh\beta,\sinh\beta\vec{n})$ is a general
point in the radial coordinate space. In the first term, this is constrained
to be related to the momentum $p^\mu$ (corresponding the test charge at the
origin) by eq.~(\ref{xmudef}), due to the delta function. In the second term
this is not the case, and using $p^\mu=m_2(1,\vec{0})$ (in Minkowski 
coordinates) one finds
\begin{equation}
j_{\tau\tau}=\frac{\kappa}{2}e^{-\tau}
\left[m_2\delta^{(3)}(\vec{x})+K\cosh\beta\right].
\label{jtautau3}
\end{equation}
That is, the effect of a constant current density for the conformally invariant
quantity $h_{\mu\tau}$ becomes a non-trivial charge density $\sim\cosh\beta$
in the current for $h_{\tau\tau}$. That this makes physical sense can be seen
as follows. We have already noted that the spurious pole of 
eq.~(\ref{gensolK2}) corresponds to a mass at $\beta=i\pi$, in addition to the
mass at $\beta=0$. Both of these masses are necessarily positive. However, the
charge density $\cosh\beta$ has the property of being {\it odd} under the
transformation
\begin{equation}
\beta\rightarrow i\pi-\beta.
\label{betatrans}
\end{equation}
Thus, the added charge density reinforces the physical charge in the upper
branch of the AdS space, but acts to cancel out the fake charge in the lower
branch, as required. Absorbing various factors in the constant charge density 
$K$, eq.~(\ref{Ktildeeq}) is modified to
\begin{equation}
\frac{1}{\sinh^2\beta}\partial_\beta\left(\sinh^2\beta\partial_\beta
\tilde{K}(\beta)\right)-3\tilde{K}(\beta)=B\cosh\beta,
\label{Ktildeeq2}
\end{equation}
whose general solution is 
\begin{equation}
\tilde{K}(\beta)=(A_1+A_3\beta)\left[\frac{1}{\sinh\beta}+2\sinh\beta\right] + A_2 \cosh\beta,
\label{gensolK3}
\end{equation}
where $A_3=B/8$ is to be determined, and $A_1$ is given by eq.~(\ref{A1sol}).
One can fix $A_3$ using the requirement that the potential must not diverge at
$\beta=i\pi$, if the spurious charge has been consistently removed:
\begin{equation}
A_3=-\frac{A_1}{i\pi}.
\label{A3sol}
\end{equation}
The solution of the Newtonian potential which respects all boundary conditions
is therefore
\begin{equation}
\Phi(\beta,\tau)=m_2 e^\tau \left[ \frac{\kappa^2}{16\pi^2}(i\beta+\pi)\left[\frac{1}
{\sinh\beta}+2\sinh\beta\right] +C \cosh\beta \right], 
\label{Phisol}
\end{equation}
where $C=A_2/m_2$ plays the same role as the constant in the QED case. 
We see that the time-dependent potential contains the combination $m_2e^\tau$,
as a consequence of the same combination occuring on the right-hand side of
eq.~(\ref{htteq3}). We can thus identify a static potential
\begin{equation}
\tilde{\Phi}(\beta)=m_2 \left[ \frac{\kappa^2}{16\pi^2}(i\beta+\pi)\left[\frac{1}
{\sinh\beta}+2\sinh\beta\right]+C \cosh\beta \right].
\label{Phisol2}
\end{equation}
This is correct at all times if we consider $m_2$ to be time-dependent
($m_2(\tau)\equiv m_2e^\tau$, where $m_2$ is the static mass), which takes 
into account the fact that gravitational Wilson lines are not invariant under 
reparametrisations of $p^\mu$. \\

Now consider adding a test particle of mass $m_1$ at location $\beta$. If $m_2$
is at $\beta=0$, then $\beta\equiv\beta_{12}$ is the geodesic separation 
between the masses in the AdS space. The total time-independent potential 
is then given by
\begin{equation}
E(\beta_{12})=m_1\tilde{\Phi}=m_1m_2\left[\frac{\kappa^2}{16\pi^2}(i\beta_{12}
+\pi)\left[\frac{1}{\sinh\beta_{12}}+2\sinh\beta_{12}\right]+C\cosh\beta_{12}
\right].
\label{Edef}
\end{equation}
Note that the potential energy is defined only up to an
arbitrary amount of the solution to the homogeneous equation\footnote{In the 
QED case, for which Laplace's equation holds, this amounts to the presence of
an arbitrary constant.}. In the present case, this includes a contribution 
proportional to $\cosh\beta$, as seen explicitly in eq.~(\ref{gensolK2}). 
The constant $C$ can be fixed, as in the QED case, by analytically continuing 
eq.~(\ref{Edef}) to the case of one incoming and one outgoing particle, 
corresponding to the transformation of eq.~(\ref{betatrans}) for the test 
particle of mass $m_1$. Then 
\begin{equation}
E(\beta_{12})\rightarrow -im_1m_2\left[\frac{\kappa^2}{16\pi^2}
\beta_{12}\left[\frac{1}{\sinh\beta_{12}}+2\sinh\beta_{12}\right]
-iC\cosh\beta_{12}\right].
\label{Edef2}
\end{equation}
This must vanish as $\beta\rightarrow0$, corresponding to the fact (in the
original Minkowski space) that the anomalous dimension vanishes for a straight
Wilson line contour with no cusp. This fixes
\begin{equation}
C=-\frac{i\kappa^2}{16\pi^2},
\label{Csol}
\end{equation}
so that the potential energy in the original setup of two final state Wilson 
lines is
\begin{equation}
E(\beta_{12})=m_1m_2\frac{\kappa^2}{16\pi^2}\left[(i\beta_{12}+
\pi)\left(\frac{1}{\sinh\beta_{12}}+2\sinh\beta_{12}\right)
-i\cosh\beta_{12}\right].
\label{Edef3}
\end{equation}
Finally, the cusp anomalous dimension in Minkowski space must be given by
\begin{equation}
\Gamma=iE=m_1m_2\frac{\kappa^2}{16\pi^2}\left[(i\pi-\beta_{12})
\left(\frac{1}{\sinh\beta_{12}}+2\sinh\beta_{12}\right)
+\cosh\beta_{12}\right].
\label{Gammares}
\end{equation}
This agrees exactly with the field theory calculation of this quantity in 
Minkowski space, which we present here in appendix~\ref{app:cuspcalc}.\\

If many different masses are present, one finds a total anomalous dimension
\begin{equation}
\Gamma_{tot}=\sum_{i<j}\frac{\kappa^2}{16\pi^2}m_im_j\left[(i\pi-\beta_{ij})
\left(\frac{1}{\sinh\beta_{ij}}+2\sinh\beta_{ij}\right)+\cosh\beta_{ij}\right],
\label{Gammatot}
\end{equation}
corresponding to the total energy in the radial coordinate space. The
sum is over all distinct pairs of outgoing particles, such that the cusp angle
in each case is given by
\begin{equation}
\cosh\beta_{ij}=-\frac{p_i\cdot p_j}{m_im_j}.
\label{betaijdef}
\end{equation}

Some further comments are in order regarding the anomalous dimension that we
have derived. Firstly, note that eq.~(\ref{Gammatot}) is not the most 
convenient way to express this result, with a view to examining the light-like
limit of $\beta_{ij}\rightarrow\infty$. 
Instead, we may pull out an overall factor of $\cosh\beta_{ij}$ in each term 
to obtain the alternative form
\begin{equation}
\Gamma=-\sum_{i<j}p_i\cdot p_j\frac{\kappa^2}{16\pi^2}\left[(i\pi-\beta_{ij})
\left(2\coth\beta_{ij}-\frac{1}{\sinh\beta_{ij}\cosh\beta_{ij}}\right)+1
\right],
\label{Gammares2}
\end{equation}
where we have used the cusp angle definition of eq.~(\ref{coshbetadef}).
This will be useful in the following section.\\

Secondly, there are potential conceptual issues regarding the gravitational
cusp anomalous dimension. This is defined in terms of the ultraviolet 
renormalisation properties of the vertex at which multiple Wilson lines meet.
In calculating any particular diagram contributing to a Wilson line expectation
value (and thus contributing to the gravitational soft function), one 
encounters additional ultraviolet singularities relating in principle to the 
renormalisation of the masses $m_i$, and the gravitational coupling constant 
$\kappa$. However, gravity is non-renormalisable, such that the latter 
contributions are ill-defined, leading to additional operators at each order
in perturbation theory that need to be included in Wilson line 
diagrams\footnote{For a recent discussion of the difficulty of considering
running parameters in quantum gravity, see~\cite{Anber:2011ut}.}. 
One may then worry that it is not possible to define a gravitational cusp 
anomalous dimension at all. However, this worry can be seen to be misguided
for the following reasons: UV singularities of Wilson lines associated with
renormalisation of the cusp correspond to IR singularities in scattering 
amplitudes, thus must be ultimately separable in a meaningful way from UV
singularities associated with Lagrangian parameters. Also, as can be inferred
from~\cite{Naculich:2011ry,White:2011yy,Akhoury:2011kq}, the gravitational cusp
anomalous dimension is {\it one-loop exact}. Thus, the situation in which one
might have to worry about separating overlapping UV divergent contributions 
to the gravitional couplings and to the multieikonal vertex (which potentially
occurs only beyond one loop order) never arises. It seems, then, that one-loop 
exactness of the gravitational anomalous dimension is crucially related to the
fact that the IR limit of perturbative general relativity must be UV finite. 

\section{Gravitational Wilson lines in the lightlike limit}
\label{sec:collinear}
In the previous section, we have obtained the cusp anomalous dimension for
perturbative GR via an energy calculation in radial coordinate space, similar
to the QED / QCD analysis of~\cite{Chien:2011wz}. To complete this analysis, 
it is instructive to examine the lightlike limit $p_i^2\rightarrow 0$.
As is well known, additional collinear singularities appear in this limit.
In gravity, these singularities are present on a diagram by diagram basis,
but cancel after summing over all diagrams~\cite{Weinberg:1965nx,
Akhoury:2011kq}. The appearance of collinear singularities in the gauge theory
case is associated in the radial coordinate space with a linear divergence of
the (imaginary) energy of two static charges as $\beta\rightarrow\infty$. A
nice comparison was outlined in~\cite{Chien:2011wz} between this phenomenon
and linear confinement - the confinement in this case being that of e.g. quarks
within jets. \\

The aim of this section is to briefly revisit this analysis in the context of
the gravity calculation carried out in the previous section, in order to 
complete the conceptual mapping between the gauge theory and gravity cases. We
begin with the form of the anomalous dimension given in eq.~(\ref{Gammares2}),
in which an explicit factor of $p_i\cdot p_j$ has been pulled out in each term.
This expression remains valid as $m_i\rightarrow 0$, in which case 
$\beta_{ij}\rightarrow\infty$ for all $j$. Let us now consider the case that
particle $i$ indeed becomes massless. From eq.~(\ref{Gammares2}), we may 
isolate all contributions involving particle $i$ as
\begin{equation}
\Gamma_i=-\sum_{j\neq i}p_i\cdot p_j\frac{\kappa^2}{16\pi^2}
\left[(i\pi-\beta_{ij})\left(2\coth\beta_{ij}-\frac{1}{\sinh\beta_{ij}
\cosh\beta_{ij}}\right)+1\right],
\label{Gammares3}
\end{equation}
which as $m_i\rightarrow 0$ ($\beta_{ij}\rightarrow\infty$) becomes
\begin{equation}
\Gamma_i\rightarrow\sum_{j\neq i}
p_i\cdot p_j\frac{\kappa^2}{8\pi^2}\beta_{ij}.
\label{Gammares4}
\end{equation}
Examining $\beta_{ij}$ itself, this is
\begin{equation}
\beta_{ij}=\cosh^{-1}\left(-\frac{p_i\cdot p_j}{m_im_j}\right)
=\log\left(-\frac{p_i\cdot p_j}{m_im_j}\right)+{\cal O}(1).
\label{coshexp}
\end{equation}
We may rewrite the logarithm on the right-hand side as
\begin{equation}
\log\left(-\frac{p_i\cdot p_j}{m_im_j}\right)=-\log\left(\frac{m_i}{Q}
\right)+\log\left(\frac{-p_i\cdot p_j}{Qm_j}\right)=
-\log\left(\frac{m_i}{Q}\right)+{\cal O}(1),
\label{coshexp2}
\end{equation}
where $Q$ is an arbitrary momentum scale to keep the arguments dimensionless.
Substituting this into eq.~(\ref{Gammares4}) gives
\begin{equation}
\Gamma_i\rightarrow-\frac{\kappa^2}{8\pi^2}\log\left(\frac{m_i}{Q}\right)
p_i\cdot\sum_{j\neq i} p_j.
\label{Gammares5}
\end{equation}
Using momentum conservation
\begin{equation}
\sum_i p_i=0,
\label{momcon}
\end{equation}
the total contribution to the cusp anomalous dimension from particle $i$ is
\begin{equation}
\Gamma_i\rightarrow\frac{\kappa^2}{16\pi^2}\log\left(\frac{m_i}{Q}\right)
p_i^2=0.
\label{Gammares6}
\end{equation}
Thus, collinear singularities do not appear in the gravitational soft function.
This is not, of course, a new result. The cancellation of soft collinear
singularities has been known since~\cite{Weinberg:1965nx}, and has been 
generalised to hard collinear singularities in~\cite{Akhoury:2011kq}. The
essential physical reason for this cancellation can also be obtained by direct
analogy with abelian and non-abelian gauge theory: collinear singularities
associated with a given particle depend on its squared charge. In QED this is
$q^2$, where $q$ is the electromagnetic charge. In QCD, this is the quadratic
Casimir invariant associated with the representation appropriate to the
parton of interest. In gravity, the squared charge is the 4-momentum squared,
which is zero if collinear singularities are to be present - which ends up 
removing them. \\

Although the above argument is formulated for only one particle becoming 
massless, it generalises straightforwardly to cases involving more than one
massless particle. Note that our reasoning does not tell us that massless
particles do not contribute at all to the energy in radial coordinate space. 
Rather, the diverging term involving the geodesic separation
of a given particle from one of its partners (as the former becomes massless) 
cancels after summing all contributions to the potential energy, from all the 
other particles. There are still terms which are ${\cal O}(\beta^0)$, which we 
neglected in the above analysis. It is in principle possible to calculate 
the total potential energy from interactions with the massless particle by 
giving the massless particle a small mass $m_i$, as above, and setting 
$m_i\rightarrow 0$ at the end of the calculation after summing all 
contributions. \\

How are we to square this with the fact that, for massive Wilson lines, we
derived the energy in the radial coordinate space after taking the Newtonian
limit? In this limit, it must be true that a massless particle contributes 
nothing to the energy of a collection of masses, as it does not gravitate. 
The resolution of this puzzle is that as $m_i\rightarrow 0$, the Newtonian
limit is no longer valid, as this relies on being able to define velocities 
which are much less than the speed of light, which is only possible for massive
particles. Instead, as $m_i\rightarrow 0$, one must take into account 
special relativistic corrections. \\

Having now completed our analysis of gravitational Wilson lines in AdS space,
and their analogies with the QED case, it is amusing to note that both theories
are in fact special cases of a general formulation, with a continuous relation
between them. This is the subject of the following section.

\section{General formulation}
\label{sec:gen}
In the previous sections, we have reviewed the properties of Wilson lines in
AdS space, and used an analogous analysis to~\cite{Chien:2011wz} to examine the
properties of the cusp anomalous dimension in perturbative GR. In this section,
we point out that one can formulate a general calculation for the cusp 
anomalous dimension, two special cases of which are QED and gravity. 
Furthermore, we will show that these cases are continuously related to each 
other. This perhaps adds an interesting additional way of thinking about the 
results of~\cite{Chien:2011wz} and the present paper, which may be of further 
use.\\

Consider the operator
\begin{equation}
{\cal W}_n=\exp\left[i\lambda\,p^{\mu_1}\,p^{\mu_2}\ldots p^{\mu_n}
\int_0^\infty ds\,
H_{\mu_1\mu_2\ldots\mu_n}(sp)\right],
\label{Wndef}
\end{equation}
which is clearly defined for integer $n$. We may recognise this as a 
generalisation of the Wilson line operators of eqs.~(\ref{Wilson}, 
\ref{Wgdef}), where $p^\mu$ is the hard momentum of a particle which emits
soft quanta of a spin-$n$ gauge field $H_{\mu_1\ldots\mu_n}$ with coupling
constant $\lambda$~\footnote{The reader may object, regarding 
how to interpret this operator for $n\geq 3$. We return to this point in what 
follows.}. We have again parametrised the straight-line contour of the Wilson
line according to eq.~(\ref{xmudef}), which allows to rewrite eq.~(\ref{Wndef})
as
\begin{align}
{\cal W}_n&=\exp\left[i\lambda\,p^{\mu_1}\,p^{\mu_2}\ldots p^{\mu_{n-1}}
\int dx^{\mu_n}\,
H_{\mu_1\ldots\mu_{n}}(x)\right]\notag\\
&=\exp\left[i\lambda\,p^{\mu_1}\,p^{\mu_2}\ldots p^{\mu_{n-1}}
\int_{-\infty}^\infty d\tau\,
H_{\mu_1\ldots\mu_{n-1}\tau}(x)\right].
\label{Wndef2}
\end{align}
We see that the Wilson line phase is the product of a reparametrisation 
invariant quantity $H_{\mu_1\ldots\mu_{n-1}\tau}$ (which has one radial-space
index and $n-1$ Minkowski-space indices) and a charge given by 
$\lambda p^{\mu_1}\ldots p^{\mu_{n-1}}$. The fact that the former is 
conformally invariant implies that 
\begin{equation}
H_{\tau\tau\ldots\tau}=\dot{x}^{\mu_1}\,\dot{x}^{\mu_2}\ldots 
\dot{x}^{\mu_{n-1}}\,H_{\mu_1\mu_2\ldots\mu_{n-1}\tau}= e^{(n-1)\tau}
\tilde{H}(\beta),
\label{Hteq}
\end{equation}
which is a generalisation of eq.~(\ref{httform}). Let us now assume that the
homogeneous equation of motion for $H_{\tau\ldots\tau}$ is
\begin{equation}
\Box H_{\tau\ldots\tau}=0
\end{equation}
(as is true for the QED and gravity cases). Implementing the behaviour of
eq.~(\ref{Hteq}) and using the covariant D'Alambertian operator of 
eq.~(\ref{dalambert}), one finds that the spatial part of $H_{\tau\ldots\tau}$
satisfies
\begin{equation}
\left[\nabla^2-(n^2-1)\right]\tilde{H}(\beta)=0.
\label{Htilde}
\end{equation}
It is straightforward to verify that this equation reduces to 
eq.~(\ref{laplace}) and eq.~(\ref{Ktildeeq}) for the QED ($n=1$) and gravity 
($n=2$) cases respectively. The general solution of the general equation is
given by 
\begin{equation}
\tilde{H}(\beta)=A_1\left(\frac{\sinh(n\beta)}{\sinh\beta}\right)
+A_2\left(\frac{\cosh(n\beta)}{\sinh\beta}\right),
\label{Htildesol}
\end{equation}
where we have chosen to write this explicitly as the sum of two parts which
have a definite parity under the transformation of eq.~(\ref{betatrans}). One
may check (using hyperbolic function identities) that this result indeed 
reproduces the QED and gravity results of eq.~(\ref{lapsol}, \ref{gensolK}). 
However, the solution of eq.~(\ref{Htildesol}) (as a function at least) is 
well-defined for any $n$. In particular, we may consider $n$, divorced from the
original context of eq.~(\ref{Wndef}), to be a continuous parameter that 
smoothly interpolates between the two solutions we obtained previously. \\

In both of the cases considered so far, the solution of eq.~(\ref{Htildesol})
did not have the right behaviour as $\beta\rightarrow\infty$ to correspond to
the cusp anomalous dimension in the relevant field theory. This was rectified
by modifying the current density in the inhomogeneous equation for the
conformally invariant field component $H_{\mu_1\ldots\mu_{n-1}\tau}$ by a 
constant density. Introducing a delta function in the radial coordinate space, 
one may write the Wilson line phase from eq.~(\ref{Wndef2}) as
\begin{equation}
i\lambda\,p^{\mu_1}\,p^{\mu_2}\ldots p^{\mu_{n-1}}
\int_{-\infty}^\infty d\tau\int d^3\vec{x}\,\delta^{(3)}(\vec{x})
H_{\mu_1\ldots\mu_{n-1}\tau}(x),
\label{jndef}
\end{equation}
such that the current density that sources the conformally invariant quantity 
$H_{\mu_1\ldots\mu_{n-1}\tau}$ is given by
\begin{equation}
\sqrt{-g}j^{\mu_1\ldots\mu_{n-1}\tau}=\lambda\,p^{\mu_1}\ldots p^{\mu_{n-1}}
\delta^{(3)}(\vec{x}),
\end{equation}
from which one finds\footnote{Again we have absorbed geometric factors into 
the delta function, so that this is normalised according to 
eq.~(\ref{deltnorm}).}
\begin{equation}
j_{\mu_1\ldots\mu_{n-1}\tau}=\lambda e^{-2\tau}\,p_{\mu_1}\,\ldots
p_{\mu_{n-1}}\delta^{(3)}(\vec{x}). 
\label{jmuteq}
\end{equation}
The appropriate generalisation of the constant charge density procedure is to
modify this current to
\begin{equation}
j_{\mu_1\ldots\mu_{n-1}\tau}=\lambda e^{-2\tau}\,p_{\mu_1}\,\ldots
p_{\mu_{n-1}}\left[\delta^{(3)}(\vec{x})+K\right], 
\label{jmuteq2}
\end{equation}
such that one finds
\begin{align}
j_{\tau\tau\ldots\tau}&=\dot{x}^{\mu_1}\ldots\dot{x}^{\mu_{n-1}}
H_{\mu_1\ldots\mu_{n-1}\tau}\notag\\
&=\lambda e^{(n-3)_\tau}m^{n-1}\left[\delta^{(3)}(\vec{x})
+K\cosh^{(n-1)}\beta\right].
\label{jteq}
\end{align}
Thus, the constant charge density becomes a smooth distribution 
$\sim\cosh^{(n-1)}\beta$ distributed throughout space. As for 
eq.~(\ref{Htildesol}), one may continue $n$ away from integer values. \\

In the QED and gravity cases, the constant charge density procedure ensured
that the energy associated with a pair of charges in AdS space diverged 
with an overall power of $\beta$ at large separations 
(corresponding to collinear singularites in 
Minkowski space). This became somewhat non-trivial in the gravity example, in
which it was crucial that the modification to $j_{\tau\tau}$ went like
$\cosh\beta$. For integer values $n\geq 3$, we may note that this property does
not generalise. One may verify that
\begin{align}
\left[\nabla^2-(n^2-1)\right]\left(\frac{(C_1\beta+C_2)\cosh(n\beta)}
{\sinh\beta}\right)&=\frac{2C_1n\sinh(n\beta)}{\sinh\beta};\label{coshn1}\\
\left[\nabla^2-(n^2-1)\right]\left(\frac{(C_1\beta+C_2)\sinh(n\beta)}
{\sinh\beta}\right)&=\frac{2C_1n\cosh(n\beta)}{\sinh\beta}.
\label{coshn2}
\end{align}
Thus, a function constructed by modifying the solution of the homogeneous
equation (eq.~(\ref{Htildesol})) to include an overall power of $\beta$, is 
not consistent with a modified charge density $\sim\cosh^{(n-1)}\beta$ in 
general (one may also show that the converse is true). Nevertheless, these 
results are correct in the QED and gravity cases. For QED, eq.~(\ref{coshn1}) 
applies, and the resulting (constant) charge density picks out the solution of
the homogeneous equation which has odd parity under the transformation of 
eq.~(\ref{betatrans}). For gravity, eq.~(\ref{coshn2}) produces the required
$\cosh\beta$ charge density, which then picks out the even solution for
$\tilde{H}(\beta)$. \\

For $n\geq 3$, it is no longer true that the charge density in 
$j_{\tau\ldots\tau}$ arising from the constant charge prescription is such as
to modify the energy by a linear term in $\beta$. This presumably means that
one cannot interpret the resulting energy as a cusp anomalous dimension of a
Wilson line operator. However, this is not at all surprising, as the operator
of eq.~(\ref{Wndef}) ceases to be meaningful for $n\geq 3$. If 
eq.~(\ref{Wndef}) is to be interpreted as the operator describing the emission
of soft higher spin gauge bosons, then Lorentz invariance demands that the
quantity
\begin{displaymath}
\sum_n g_n\, p^{\mu_1}\ldots p^{\mu_{n-1}}
\end{displaymath}
be conserved, where the sum is over all external particles $n$, and $g_n$ is a
constant which may depend on a given hard particle in general 
(see e.g. chapter 13 of~\cite{Weinberg:1995mt}). The $n=1$ and $n=2$ cases
correspond to electromagnetic charge and 4-momentum conservation respectively.
However, for $n\geq 3$ no conserved quantity is possible if non-trivial
scattering is to occur, which makes the above argument meaningless if 
$n\geq 3$ (for integer $n$). That this is seen directly in the general 
analysis, in terms of the charge densities not matching up, is interesting.\\

In this section, we have considered a general formulation of the cusp anomalous
dimension calculation in Minkowski space, in which the QED and gravity theories
emerge as special cases. Indeed, as discussed above, the general Wilson line 
operator of eq.~(\ref{Wndef}) is only meaningful (for integer $n$) for $n=1$, 
2. These are precisely the cases of QED and gravity respectively. Nevertheless,
it is amusing to note that one may regard $n$ also as a continuous parameter
which smoothly interpolates between the one loop cusp anomalous dimensions
of QED and GR. Our main motivation in pointing this out is in case this has
any further interpretation or utility, in addition to novelty. 

\section{Conclusion}
\label{sec:conclusion}
In this paper, we have examined gravitational Wilson lines, representing the 
emission of soft gravitons from an eikonal emitter. We have examined in detail
the anomalous dimension which controls the renormalisation of the vertex in a 
general correlator of such Wilson lines, with a view to generalising the 
analysis of~\cite{Chien:2011wz} which relates the calculation of such anomalous
dimensions to static energies in Euclidean AdS space. \\

There are a number of motivations for doing this. To start with, gravitational
Wilson line operators and their renormalisation properties have only recently
been studied. Given that the cusp anomalous dimension forms such a crucial 
object in gauge theories, any investigation of its properties in a 
gravitational context is interesting enough by itself. Our analysis also sheds
further light on the radial coordinate picture for gauge theory Wilson lines
developed in~\cite{Chien:2011wz}, due to a number of subtleties in
the gravitational case. For example, the procedure of adding a constant charge
density to obtain the correct boundary conditions in the QED analysis implies
a $\cosh\beta$ charge density in the gravity case, which is precisely such as
to lead to the correct solution for the anomalous dimension, and also to be 
consistent with the fact that masses, unlike electric charges, can only be 
positive. \\

The similarities and differences between the gravity and QED / QCD radial 
coordinate pictures may be useful in providing further insight and intuition 
in current research at the boundary between gauge and gravity theories. 
One example of this is the property of jet confinement in QCD, which has a 
particularly intuitive description in the radial coordinate picture, and which
is absent in gravity. Another example where the radial coordinate picture might
be useful is in thinking about the all-order structure of IR singularities. 
Radial coordinates were used in~\cite{Chien:2011wz} to motivate a family of
{\it conformal gauges}, in which the contributions from graphs involving 
multiple gluon vertices was much reduced (e.g. absent at two loop level). It
seems likely that this is related to the conjectured dipole structure of 
soft singularities in QCD~\cite{Gardi:2009qi,Dixon:2009ur,Becher:2009cu,
Becher:2009qa}. In gravity there is less immediate motivation for considering
such conformal gauges, due to the fact that the Wilson line operator is not
conformally invariant, and also that the cusp anomalous dimension is known to
be one loop exact. However, it may be that the simple structure of IR
divergences in gravity (which is manifestly of dipole form), and the 
relationship between the radial coordinate pictures in both the gravity and
gauge theory cases, can be used to gain further insight into the dipole 
conjecture.\\

We also considered a general formulation of the cusp anomalous dimension
calculation, in which a smooth parameter arises which smoothly interpolates
between the QED and gravity cases. This analysis breaks down for $n\geq 3$
(where $n$ is the number of momentum factors in the Wilson line operator), but
this is entirely consistent with the fact that such operators are not 
physically meaningful due the fact that they lead to conserved higher-rank
tensorial charges and thus trivial scattering. It is not immediately obvious
what the applications of $n$ considered as a continuous parameter might be, 
but one might hope that it may lead to further insight on the relationship 
between QED (or its non-Abelian brother, QCD) and gravity. \\

In summary, the radial coordinate picture potentially offers novel new insights
into both gauge and gravity theories, and the relationships between them. The
results of this paper provide a useful addition in this regard.

\section*{Acknowlegments}
We thank Rachel Dowdall, Einan Gardi, Mark Harley and Jack Laiho for 
discussions and / or comments on the manuscript. CDW is supported by the STFC 
Postdoctoral Fellowship ``Collider Physics at the LHC''. DJM acknowledges 
partial support from the STFC Consolidated Grant ST/G00059X/1.

%%%%%%%%%%%%%%%%

\appendix
\section{Newton's law from the gravitational Wilson line}
\label{app:newton}
\begin{figure}
\begin{center}
\scalebox{0.8}{\includegraphics{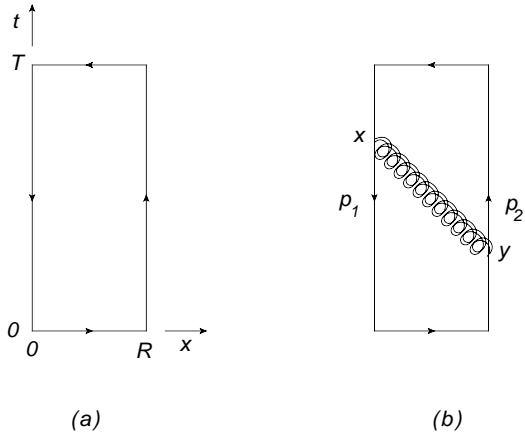}}
\caption{(a) Wilson loop contour in Minkowski space used for the calculation
of the Newtonian potential; (b) the relevant one loop diagram.}
\label{fig:contour}
\end{center}
\end{figure}

In this appendix, we show how Newton's law of gravity can be calculated using
the Wilson line operator of eq.~(\ref{Wgdef}). First, we consider the Minkowski
space contour ${\cal C}$ of figure~\ref{fig:contour}(a). 
It is a textbook result in field theory (see e.g.~\cite{Montvay:1994cy}) that 
the static potential $V(R)$ between two charges of separation $R$ is given by
\begin{equation}
V(R)=\lim_{T\rightarrow\infty}\frac{i}{T}\log\langle{\cal W}({\cal C})\rangle,
\label{VRdef}
\end{equation}
where $\langle {\cal W}({\cal C})\rangle$ is the vacuum expectation value of 
the Wilson loop along the contour ${\cal C}$, and the latter is given in the 
present case by
\begin{equation}
{\cal W}({\cal C})=\exp\left[i\frac{\kappa}{2}\oint_{\cal C} ds\,p^\mu\,
p^\nu\,h_{\mu\nu}(sp^\mu)\right].
\label{Wdefapp}
\end{equation}
As the time $T\rightarrow\infty$, the only contributing diagram at one loop 
order is that shown in 
figure~\ref{fig:contour}(b), where the gravitons are emitted at positions 
\begin{equation}
x=sp_1,\quad y=tp_2,
\label{xydef}
\end{equation}
and the 4-momenta of the static masses are given by
\begin{equation}
p_1=(m_1,\vec{0}), \quad p_2=(m_2,\vec{0}).
\end{equation}
One may evaluate this using the position space graviton propagator (in four 
dimensions)
\begin{equation}
D_{\mu\nu,\alpha\beta}(x-y)=\frac{1}{4\pi^2}\frac{P_{\mu\nu,\alpha\beta}}
{(x-y)^2-i\epsilon},
\label{proppos}
\end{equation}
as can be obtained by Fourier transforming the momentum space propagator of
eq.~(\ref{prop}). The diagram of figure~\ref{fig:contour}(b) then gives a
contribution 
\begin{equation}
\log{\cal W}_g=\left(\frac{\kappa}{2}\right)^2\frac{1}{4\pi^2}
\int_{T/m_1}^0 ds\int_0^{T/m_2}dt\frac{P_{\mu\nu,\alpha\beta}\,p_1^\mu\,
p_1^\nu\,p_2^\alpha\,p_2^\beta}{(x-y)^2-i\epsilon}+{\cal O}(\kappa^4),
\label{F1}
\end{equation}
where we have used the fact that the one-loop contribution to the Wilson loop
expectation value is the same as the contribution to the exponent at this 
order\footnote{In fact, the one-loop diagram enters the exponent to all 
orders, due to the gravitational analogue of Abelian 
exponentiation~\cite{Weinberg:1965nx,Naculich:2011ry,White:2011yy,
Akhoury:2011kq}.}. Using the definition of $P_{\mu\nu,\alpha\beta}$ from
eq.~(\ref{prop}), one finds
\begin{equation}
P_{\mu\nu,\alpha\beta}\,p_1^\mu\,p_1^\nu\,p_2^\alpha\,p_2^\beta=
2(p_1\cdot p_2)^2-p_1^2p_2^2=m_1^2m_2^2.
\label{Pp}
\end{equation}
Also transforming the integrals in eq~(\ref{F1}) to $x^0=sm_1$ and $y^0=sm_2$,
one finds
\begin{align}
\log{\cal W}_g&=\frac{\kappa^2m_1m_2}{16\pi^2}\int_T^0dx^0\int_0^Tdy_0
\frac{1}{[-(x^0-y^0)^2+R^2-i\epsilon]}\notag\\
&\simeq\frac{\kappa^2m_1m_2T}{16\pi^2}\int_{-T}^Tdy_0
\frac{1}{(y^0)^2-R^2+i\epsilon},
\label{F2}
\end{align}
where in the second line we have used the fact that we are taking 
$T\rightarrow\infty$. The $y^0$ integral gives
\begin{align}
\int_{-T}^{T}dy_0\frac{1}{(y^0)^2-R^2+i\epsilon}&\simeq\int_{-\infty}^{\infty}
dy_0\frac{1}{(y^0-R+i\epsilon)(y^0+R-i\epsilon)}\notag\\
&=2\pi i\left(-\frac{1}{2R}\right),
\label{y0int}
\end{align}
using Cauchy's theorem. Finally one finds
\begin{equation}
\log{\cal W}_g=-\frac{i\kappa^2Tm_1m_2}{16\pi R}.
\label{Wgres}
\end{equation}
Substituting this result into eq.~(\ref{VRdef}) and using the definition of
$\kappa$ in terms of Newton's constant, $\kappa=\sqrt{16\pi G_N}$, gives
\begin{equation}
V(R)=\frac{G_Nm_1m_2}{R}.
\label{VRres}
\end{equation}
The force between the two charges is thus
\begin{equation}
F=-\frac{G_Nm_1m_2}{R^2},
\label{Fdef}
\end{equation}
which is Newton's law of gravity.\\

Note that this calculation is closely related to an analysis carried out 
in~\cite{Hamber:1994jh}, which studies nonperturbative quantum corrections to
the Newtonian potential. To this aim, the authors define a Wilson line operator
obtained by integrating over the worldline of a particle of mass $m$:
\begin{equation}
W_{\rm wl}=\exp\left[-im\int_C dt \sqrt{-g_{\mu\nu}\frac{dx^\mu}{dt}
\frac{dx^\nu}{dt}}\right],
\label{Wwldef}
\end{equation}
where $C$ is the worldline contour, and $t$ the proper time along this
path. This operator is equivalent, in the weak field limit, to the operator
of eq.~(\ref{Wgdef}), as we now show. The weak field expansion of 
eq.~(\ref{gmunuexp}) allows us to rewrite eq.~(\ref{Wwldef}) as
\begin{equation}
W_{\rm wl}=\exp\left[-i\int_C dt \sqrt{-\left(\eta_{\mu\nu}+\kappa h_{\mu\nu}
\right)p^\mu\,p^\nu}\right],
\label{Wwl2}
\end{equation}
where we have also taken the mass factor inside the square root and used 
$p^\mu=mdx^\mu/dt$. Expanding the square root to first order in the graviton
field gives
\begin{equation}
W_{\rm wl}=\exp\left[-im\int_C dt +i\frac{\kappa}{2}\int_C dsh_{\mu\nu}
p^\mu\,p^\nu
\right],\quad s=\frac{t}{m}.
\label{Wwl3}
\end{equation}
The first term in the exponent is absorbed into the normalisation of the Wilson
line operator. The second term is precisely that of eq.~(\ref{Wgdef}). That 
this must be the case follows from the fact that in the eikonal approximation,
a particle emitting soft gravitons does not recoil, and thus
follows its classical trajectory. Its action must then be given by its 
classical action, which is indeed the integral over the worldline as in 
eq.~(\ref{Wwldef}). This then fixes the form of the interaction between the
eikonal particle and the (soft) graviton field.

\section{Normalisation of the solution to Newton's equation}
\label{app:norm}
\begin{figure}
\begin{center}
\scalebox{1.0}{\includegraphics{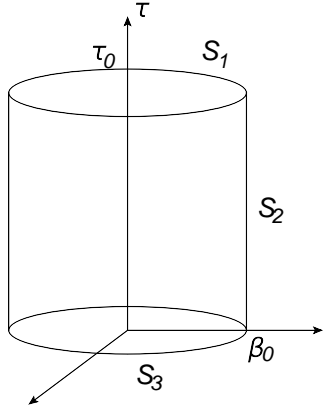}}
\caption{The 3-surface used to normalise the coefficient $A_1$ in the general
solution to the homogeneous Newton equation. The horizontal plane is 
representative of the spacelike directions, where the radial distance in this
plane represents $\beta$.}
\label{fig:surface}
\end{center}
\end{figure}

In section~\ref{sec:wilson}, we construct eq.~(\ref{htteq3}) for the Newtonian
potential $\Phi$, and show that the general solution of the homogeneous 
equation is given by eq.~(\ref{gensolK}). In this appendix, we show how the
constant $A_1$ can be related to the strength of the source term occuring on 
the right-hand side of eq.~(\ref{htteq3}). First, note that rewriting this 
equation in terms of the covariant d'Alambertian operator gives
\begin{equation}
\frac{1}{\sqrt{-g}}\partial_\mu\left(\sqrt{-g}g^{\mu\nu}\partial_\nu \Phi
\right)=\left(\frac{\kappa}{2}\right)^2m_2e^{-\tau}\delta^{(3)}(\vec{x}).
\label{newtonlaw2}
\end{equation}
Integrating over the complete 4-volume on both sides and using the covariant 
form of the divergence theorem~\cite{Landau}, one finds
\begin{equation}
\int_S dS_\mu \,\sqrt{-g}\,g^{\mu\nu}\,\partial_\nu\Phi=
\left(\frac{\kappa}{2}\right)^2\int d\Omega\,\sqrt{-g}
\,m_2\,e^{-\tau}\,\delta^{(3)}(\vec{x}),
\label{surfint1}
\end{equation}
where $\Omega$ is the 4-volume, and $dS_\alpha$ the element of 3-surface area.
The right-hand side gives
\begin{equation}
\left(\frac{\kappa}{2}\right)^2
\int d\Omega\,\sqrt{-g}\,m_2\,e^{-\tau}\,\delta^{(3)}(\vec{x})
=\left(\frac{\kappa}{2}\right)^2 m_2\int d\tau e^{3\tau},
\label{surfintRHS}
\end{equation}
where we have used the fact that $\sqrt{-g}\sim e^{4\tau}$, and cancelled
spacelike geometric factors in $\sqrt{-g}$ with similar factors in 
$\delta^{(3)}(\vec{x})$, consistent with our definition of the delta function 
in eq.~(\ref{deltnorm}). We now consider the surface shown in 
figure~\ref{fig:surface}, and consisting of a cylinder of radius $\beta_0$ in
the spacelike direction, with upper and lower surfaces at $\tau=\tau_0$ and
$\tau=0$ respectively.  
Then eq.~(\ref{surfintRHS}) becomes
\begin{align}
\left(\frac{\kappa}{2}\right)^2
\int d\Omega\,\sqrt{-g}\,m_2\,e^{-\tau}\,\delta^{(3)}(\vec{x})
&=\left(\frac{\kappa}{2}\right)^2 m_2\int_0^{\tau_0} d\tau e^{3\tau}
\notag\\
&=\left(\frac{\kappa}{2}\right)^2\frac{1}{3} m_2(e^{3\tau_0}-1).
\label{surfintRHS2}
\end{align}
Splitting up the surface into parts as labelled in figure~\ref{fig:surface}, 
the surface integrals over $S_1$, $S_2$ and $S_3$ are given by
\begin{align}
\int_{S_1} dS_\mu \,\sqrt{-g}\,g^{\mu\nu}\,\partial_\nu\Phi
&=-4\pi\,e^{3\tau_0}\int_0^{\beta_0} d\beta\sinh^2\beta\tilde{K}(\beta);
\label{intS1}\\
\int_{S_2} dS_\mu \,\sqrt{-g}\,g^{\mu\nu}\,\partial_\nu\Phi
&=\frac{4\pi}{3}\sinh^2\beta_0\partial_\beta\,\left[\tilde{K}(\beta)
\right]_{\beta_0}(e^{3\tau_0}-1);\label{intS2}\\
\int_{S_3} dS_\mu \,\sqrt{-g}\,g^{\mu\nu}\,\partial_\nu\Phi
&=4\pi\int_0^{\beta_0} d\beta\sinh^2\beta\tilde{K}(\beta),
\label{intS3}
\end{align}
where we have used eq.~(\ref{Ktildedef}). Substituting explicitly the general
solution of eq.~(\ref{gensolK}) gives a total surface integral of
\begin{equation}
\int_{S} dS_\mu \,\sqrt{-g}\,g^{\mu\nu}\,\partial_\nu\Phi
=\frac{4\pi}{3} A_1(e^{3\tau_0}-1),
\label{surfint3}
\end{equation}
such that equating this with eq.~(\ref{surfintRHS2}) gives
\begin{equation}
A_1=\frac{\kappa^2}{16\pi}m_2,
\label{A1sol}
\end{equation}
as has been used in eq.~(\ref{gensolK2}).

\section{Field theory calculation of the gravitational cusp anomalous 
dimension}
\label{app:cuspcalc}
\begin{figure}
\begin{center}
\scalebox{1.0}{\includegraphics{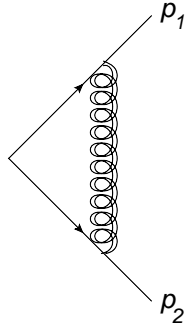}}
\caption{Diagram entering the calculation of the cusp anomalous dimension.}
\label{fig:cusp}
\end{center}
\end{figure}

In this appendix, we detail the calculation of the gravitational cusp anomalous
dimension in Minkowski space, using a conventional field theory calculation.
First, we consider the diagram shown in figure~\ref{fig:cusp}, consisting
of a graviton exchange between the two contours on either side of the cusp.
We consider the case that $p_1$ and $p_2$ are both outgoing, and correspond to
different masses $m_1$ and $m_2$. Thus, figure~\ref{fig:cusp} is not a cusp in
the traditional sense, but rather can be generally embedded into a graph where
multiple contours intersect. Using the position space propagator of 
eq.~(\ref{proppos}), this gives a contribution
\begin{equation}
{\cal F}=\left(-\frac{\kappa}{2}\right)^2\frac{1}{4\pi^2}p_1^\mu\,p_1^\nu\,
P_{\mu\nu,\alpha\beta}\,p_2^\alpha\,p_2^\beta\int_0^\infty ds\int_0^\infty dt\,
\frac{1}{(sp_1-tp_2)^2}.
\label{Fadef}
\end{equation}
where the tensor $P_{\mu\nu,\alpha\beta}$ is defined in eq.~(\ref{prop}).
In the (--,+,+,+) metric we are using for gravitational calculations, the cusp
angle is given by
\begin{equation}
\cosh\beta_{12}=-\frac{p_1\cdot p_2}{m_1m_2},
\label{coshbetadef}
\end{equation}
where $p_i^2=-m_i^2$. Then eq.~(\ref{Fadef}) can be written, after transforming
$s\rightarrow s/m_1$, $t\rightarrow t/m_2$,
\begin{equation}
{\cal F}=-\left(-\frac{\kappa}{2}\right)^2\frac{1}{4\pi^2}\frac{p_1^\mu\,
p_1^\nu\,P_{\mu\nu,\alpha\beta}\,p_2^\alpha\,p_2^\beta}{m_1m_2}
\int_0^\infty ds\int_0^\infty dt\,\frac{1}{s^2+t^2-2st\cosh\beta_{12}}.
\label{Fadef2}
\end{equation}
The integral over $s$ and $t$ is common to the QED case. However, we evaluate 
this here for completeness. First one sets $s=\lambda t$ to give
\begin{equation}
\int_0^\infty ds\int_0^\infty dt\,\frac{1}{s^2+t^2-2st\cosh\beta_{12}}
=\int_0^\infty \frac{dt}{t}\int_0^\infty d\lambda\frac{1}
{1+\lambda^2-2\cosh\beta_{12}}.
\label{intst}
\end{equation}
The $t$ integral contains an ultraviolet and infrared divergence, where the 
coefficient of the former gives the contribution to the cusp anomalous 
dimension at one loop order. At a given loop order, care must be taken to 
consistently separate the IR and UV singular parts, which becomes especially 
cumbersome when collinear singularities are present. At one loop, however, we 
may simply evaluate the $t$ integral by imposing UV and IR cutoffs (a similar 
procedure is used in the QED analysis of~\cite{Chien:2011wz}). That is, we may
write
\begin{equation}
\int_0^\infty\frac{dt}{t}\rightarrow\int_{\Lambda_{\rm UV}}^{\Lambda_{\rm IR}}
\frac{dt}{t}=\log\left(\frac{\Lambda_{\rm IR}}{\Lambda_{\rm UV}}\right).
\label{tint}
\end{equation}
Then one finds
\begin{equation}
\int_0^\infty ds\int_0^\infty dt\,\frac{1}{s^2+t^2-2st\cosh\beta_{12}}
=-\log\left(\frac{\Lambda_{\rm UV}}{\Lambda_{\rm IR}}\right)
\int_{-\cosh\beta_{12}}^\infty d\lambda\frac{1}{\lambda^2-\sinh^2\beta_{12}},
\label{intst2}
\end{equation}
where we have also completed the square in the $\lambda$ integral, and 
transformed $\lambda\rightarrow\lambda-\cosh\beta_{12}$. Substituting 
$\lambda=\sinh\beta_{12}\,\coth u$, the $\lambda$ integral can be carried out 
to give\footnote{Care is needed with the minus sign in this equation, where the
$i\pi$ results from correctly implementing the $i\epsilon$ prescription in the
graviton propagator. Alternatively, one may carry out the calculation for one
momentum incoming and one outgoing, before analytically continuing 
$\beta_{12}=i\pi-\beta_{12}$.}
\begin{equation}
\int_{-\cosh\beta_{12}}^\infty d\lambda\frac{1}{\lambda^2-\sinh^2\beta_{12}}
=\frac{i\pi-\beta_{12}}{\sinh\beta_{12}}.
\label{lambdaint}
\end{equation}
Substituting eqs.~(\ref{intst2}, \ref{lambdaint}) into eq.~(\ref{Fadef2}) and
taking the coefficient of $\log\Lambda_{\rm UV}$, one finds a contribution
to the cusp anomalous dimension given by
\begin{equation}
\Gamma=\frac{\kappa^2}{16\pi^2}\frac{p_1^\mu\,
p_1^\nu\,P_{\mu\nu,\alpha\beta}\,p_2^\alpha\,p_2^\beta}{m_1m_2}\frac{i\pi
-\beta_{12}}{\sinh\beta_{12}}.
\label{cuspa}
\end{equation}
The kinematic factor is
\begin{align}
\frac{p_1^\nu\,P_{\mu\nu,\alpha\beta_{12}}\,p_2^\alpha\,p_2^\beta}{m_1m_2}
&=m_1\,m_2\left[\frac{2(p_1\cdot p_2)^2-m_1^2m_2^2}{m_1^2m_2^2}\right]\notag\\
&=m_1\,m_2\left[2\cosh^2\beta_{12}-1\right]\notag\\
&=m_1m_2\left[1+2\sinh^2\beta_{12}\right].
\label{kinfac}
\end{align}
Finally, one has
\begin{equation}
\Gamma=\frac{\kappa^2}{16\pi^2}m_1\,m_2(i\pi-\beta_{12})\left[
\frac{1}{\sinh\beta_{12}}+2\sinh\beta_{12}\right].
\label{cuspa2}
\end{equation}
This is not the whole story. One must also add self-energy diagrams associated
with each external line. Rather than calculate these directly, one can surmise
their contribution as follows. Firstly, the effect of 
each self-energy diagram can only depend upon the quantum numbers of a single
parton leg, so that the sum over all self-energy contributions has the form
\begin{equation}
\sum_{i=1}^L C p_i^2=\sum_{i=1}C\left[\left(\sum_{i=1}^L p_i\right)^2
-2\sum_{j>i}p_i\cdot p_j\right],
\label{selfenergy}
\end{equation}
where $C$ is a constant independent of the parton index $i$, and we have 
rewritten the momentum dependence on the right-hand side. Using momentum 
conservation (eq.~(\ref{momcon})), one may rewrite eq.~(\ref{selfenergy}) as
\begin{equation}
\sum_{i=1}^L C p_i^2=-2C\sum_i\sum_{j>i}p_i\cdot p_j,
\label{selfenergy2}
\end{equation}
and one sees that each pair of external lines is associated with the 
contribution
\begin{align}
-2Cp_i\cdot p_j&=2C\,m_i\,m_j\left(-\frac{p_i\cdot p_j}{m_i\,m_j}\right)\notag\\
&=2C\,m_i\,m_j\,\cosh\beta_{ij},
\label{selfenergy3}
\end{align}
where we have used the cusp angle definition of eq.~(\ref{betaijdef}).
Adding this to eq.~(\ref{cuspa2}), one may fix the constant $C$ by 
requiring that the cusp anomalous dimension vanishes at $\beta_{12}=i\pi$, 
corresponding to a straight line
Wilson contour with no cusp. The complete result is then
\begin{equation}
\Gamma=\frac{\kappa^2}{16\pi^2}m_1\,m_2\left[(i\pi-\beta_{12})\left(
\frac{1}{\sinh\beta_{12}}+2\sinh\beta_{12}\right)+\cosh(\beta_{12})\right].
\label{cuspa3}
\end{equation}
Note that in the large $\beta$ limit, this has the behaviour
\begin{equation}
\Gamma(\beta)\sim \beta e^\beta,
\label{largebeta}
\end{equation}
as discussed in section~\ref{sec:wilson}.

\bibliography{refs.bib}
\end{document}